\documentclass[twocolumn,floatfix,pra,showpacs,preprintnumbers,superscriptaddress,amsmath,amssymb,10pt,aps]{revtex4}

\usepackage{mathptmx}
\usepackage{subfigure}
\usepackage{psfrag,graphicx}
\usepackage{dcolumn}
\usepackage{amsmath,amssymb}
\usepackage{bm}
\usepackage{color}
\usepackage{latexsym}
\usepackage{epstopdf}
\usepackage{color}
\usepackage[english]{babel}
\usepackage{latexsym}
\usepackage{psfrag,graphicx}
\usepackage{subfigure}
\usepackage{amsmath}
\usepackage{amssymb}
\usepackage{amsfonts}
\usepackage{bm}
\usepackage{natbib}
\usepackage{epstopdf}
\usepackage{appendix}

\definecolor{mygrey}{gray}{0.35}
\definecolor{myblue}{rgb}{0.2,0.2,0.8}
\definecolor{myzard}{cmyk}{0,0,0.05,0}
\definecolor{mywhite}{rgb}{1,1,1}
\definecolor{mywhite}{rgb}{1,1,1}
\definecolor{myred}{rgb}{1,0.,0.3}

\usepackage[colorlinks=true,citecolor=myblue,linkcolor=myred]{hyperref}

\def\ba{\begin{align}}
\def\enda{\end{align}}
\def\bi{\begin{itemize}}
\def\ei{\end{itemize}}

\def\be{\begin{equation}}
\def\ee{\end{equation}}
\def\bea{\begin{eqnarray}}
\def\eea{\end{eqnarray}}
\def\bse{\begin{subequations}}
\def\ese{\end{subequations}}

\begin{document}
\title{Laser-free method for creation of two-mode squeezed state and beam-splitter transformation with trapped ions}
\author{Bogomila S. Nikolova}
\affiliation{Department of Physics, St. Kliment Ohridski University of Sofia, James Bourchier 5 blvd, 1164 Sofia, Bulgaria}
\author{Peter A. Ivanov}
\affiliation{Department of Physics, St. Kliment Ohridski University of Sofia, James Bourchier 5 blvd, 1164 Sofia, Bulgaria}

\begin{abstract}
We propose a laser-free method for creation of a phonon two-mode squeezed state and a beam-splitter transformation, using time-varying electric fields and non-linear couplings between the normal modes in a linear ion crystal. Such non-linear Coulomb-mediated interactions between the collective vibrational modes arise under specific trap-frequency conditions in an ion trap. We study the quantum metrological capability for parameter estimation of the two quantum states and show that a Heisenberg limit of precision can be achieved when the initial state with $n$ phonons evolves under the action of the beam-splitter transformation. Furthermore, we show that the phonon non-linearity and the spin-dependent force can be used for creation of a three-qubit Fredkin gate.

\end{abstract}

%\pacs{
%03.67.Ac, %Quantum computation architectures and implementations
%03.67.Bg,
%03.67.Lx,
%42.50.Dv %Coherent control of atomic interactions with photons
%}
\maketitle

%%%%%%%%%%%%%%%%%%%%%%%%%%%%%%%%%%%%%%%%%%%%%%%%%%%%%%%%%%%%%%%%%%%%%%%%%%%
%%%%%%%%%%%%%%%%%%%%%%%%%%%%%%%%%%%%%%%%%%%%%%%%%%%%%%%%%%%%%%%%%%%%%%%%%%%
%%%%%%%%%%%%%%%%%%%%%%%%%%%%%%%%%%%%%%%%%%%%%%%%%%%%%%%%%%%%%%%%%%%%%%%%%%%
%========================================================================
%========================================================================
\section{Introduction}

One of the most promising quantum platforms with applications in the realization of elementary quantum processors and high-precision quantum metrology is the system of laser-cooled trapped ions \cite{Blatt2008,Leibfried2003}. The balance between the Coulomb repulsion and the harmonic confinement leads to the formation of an ion crystal, where the ions undergo small oscillations around their equilibrium positions. Within the harmonic approximation, the ion's vibration can be expressed in terms of uncoupled collective vibrational modes \cite{James1998}. However, under a given resonance trap-frequency condition, the higher-order terms, that arise due to the Coulomb interaction, cannot be neglected, which causes coupling between the different collective vibrational modes \cite{Marquet2003}. Such a non-linear cross-Kerr coupling between the collective modes was experimentally observed in a system of two trapped ions \cite{Roos2008,Nie2009}. Recently, a quantum simulation of a trilinear Hamiltonian was demonstrated with two and three trapped ions \cite{Ding2018,Maslennikov,Ding2017,Ding2017_1}.

Here, we propose a \emph{laser-free} method for the creation of two-mode squeezed state by using non-linear interaction between the vibrational modes, in which one phonon in the axial $a$-mode is converted into one phonon in the radial $b$- and $c$-modes, and vice versa. We show that, by applying a time-varying electric field on $a$-mode, the energy exchange between the latter mode and the other two radial modes can be suppressed, which causes two-mode squeezing between $b$- and $c$- modes. The essence of this technique is that it relies only on using electric field, instead of laser fields with specific laser frequency conditions \cite{Cardoso2021}, which thus simplifies the physical realization. Furthermore, we show that the same technique can be used for the creation of a beam-splitter transformation. In that case the time-varying electric field is applied on the $b$-mode and causes an effective beam-splitter transformation between $a$- and $c$-modes.

We discuss the metrological properties of the two-mode squeezed state and the beam-splitter transformation. Recently, a parameter estimation of the non-linear coupling was studied in \cite{Ivanov2022}. Here, we show that the magnitude of the two-mode squeezed state can be estimated by measuring the phonon state probabilities. We show that, using beam-splitter transformation, one can significantly enhance the parameter estimation to a sub-Heisenberg limit of precision. In that case, the initial state, which contains $n$ phonons in each mode, evolves unitarily under the action of the beam-splitter transformation. We quantify the estimation precision in terms of quantum Fisher information (QFI). We show that the measurement uncertainty can reach a sub-Heisenberg limit of sensitivity, where the QFI scales as $n^{2}+n$. We discuss the optimal measurement basis in which the quantum Cram\'er-Rao bound is saturated, and show that the measurement of the phonon state probabilities of one of the vibrational modes leads to equality between the classical and quantum Fisher information.

Finally, we extend the discussion to the creation of conditional beam-splitter transformation. In that case a spin-dependent force acts either on the radial $b$- or the radial $c$- mode. As an example, we propose physical realization of a quantum Fredkin gate. Such a three-qubit quantum gate was recently experimentally implemented with trapped ions \cite{Gan2020}. We show that the phonon non-linear coupling and the spin-dependent force can be used for the realization of a Fredkin gate, where the control qubit is composed of the internal states of the trapped ion, and the target qubits are the Fock states of the $a$- and $c$- vibrational modes. We show that our method can be used for the creation of high-fidelity spin-dependent N00N states.

The paper is organized as follows: In Section \ref{NLPI} we provide theoretical framework for non-linear phonon-phonon interaction that can arise due to the higher-order terms in the expansion of the Coulomb potential. In Sections \ref{TM} and \ref{BS} we propose physical realization of a two-mode squeezed state and a beam-splitter transformation based on the combination of time-varying electric fields and phonon-phonon coupling. The quantum metrological properties of the two-mode squeezed state and the beam-splitter transformation are discussed in Sec. \ref{QM}. We show that, for initial state with $n$ phonons in each mode, one can achieve a sub-Heisenberg limit of precision. In Sec. \ref{CBST} the realization of a quantum Fredkin gate, based on a spin-dependent force and non-linear phonon interaction is discussed. Finally, the conclusions are presented in Sec. \ref{C}.

\section{Non-linear phonon-phonon interaction}\label{NLPI}
Here, we briefly discuss the intrinsic motional non-linearity which arises due to the long-range Coulomb interaction between the trapped ions. We consider a system of $N$ ions with electric charge $e$ and mass $m$, confined in a linear Paul trap. The total potential energy of the system includes the harmonic confinement and mutual Coulomb repulsion. It reads \cite{Leibfried2003,James1998,Marquet2003}
\begin{equation}
\hat{V}=\frac{m}{2}\sum_{\chi=x,y,z}\sum_{l=1}^{N}\omega_{\chi}^{2}\hat{r}^{2}_{\chi,l}+\sum_{l>k}\frac{e^{2}}{4\pi\epsilon_{0}|\hat{\textbf{r}}_{l}-\hat{\textbf{r}}_{k}|}.
\end{equation}
Here $\hat{\textbf{r}}_{l}=(\hat{r}_{x,l},\hat{r}_{y,l},\hat{r}_{z,l})$ is the position vector for ion $l$ and $\epsilon_{0}$ is the permittivity of free space. The trap frequencies are $\omega_{\chi}$ ($\chi=x,y,z$), where we assume that the transverse trap frequencies $\omega_{x,y}$ are much larger than the axial trap frequency $\omega_{z}$. This condition ensures that each ion occupies equilibrium position $\hat{\textbf{r}}^{(0)}_{l}=(0,0,z^{(0)}_{l})$ along the $z$ trap axis. The small displacement around an ion's equilibrium position is denoted by $\delta \hat{r}_{\chi,l}$. Within the harmonic approximation, the Hamiltonian for the ion's vibration is given by
\begin{equation}
\hat{H}_{0}=\sum_{\chi=x,y,z}\sum_{n=1}^{N}\left\{\frac{\hat{P}^{2}_{\chi,n}}{2m}+\frac{m\omega^{2}_{\chi,n}}{2}\hat{Q}^{2}_{\chi,n}\right\},\label{H0}
\end{equation}
where $\omega_{\chi,n}$ is the collective vibration frequency along the spatial direction $\chi$. Here we have introduced the normal-mode coordinate operators $\hat{Q}_{\chi,n}=\sum_{l=1}^{N}M^{\chi}_{l,n}\delta \hat{r}_{\chi,l}$ and the corresponding conjugate momentum operators $\hat{P}_{\chi,n}=\sum_{l=1}^{N}M^{\chi}_{l,n}\hat{p}_{\chi,l}$, with $M^{\chi}_{l,n}$ being the amplitude of the normal mode $n$ on ion $l$. The Hamiltonian (\ref{H0}) indicates that, up to lowest-order harmonic approximation, the vibrational modes are decoupled.

Higher-order terms in the Taylor-expansion of the Coulomb potential give rise to the coupling between different vibration modes. The non-linear Hamiltonian reads \cite{Marquet2003},
\begin{eqnarray}
\hat{H}_{\rm nl}&=&\frac{M\omega^{2}_{z}}{l}\sum_{p,m,n=1}^{N}D_{mnp}\hat{Q}_{z,p}(2\hat{Q}_{z,m}\hat{Q}_{z,n}-3\hat{Q}_{x,m}\hat{Q}_{x,n}\notag\\
&&-3\hat{Q}_{y,m}\hat{Q}_{y,n})+O(\hat{Q}^{4}_{\chi,p}),
\end{eqnarray}
where $D_{p,m,n}$ are the coupling coefficients which depend on the number of ions.  It is convenient to introduce a phonon creation and annihilation operators $\hat{a}^{\dag}_{\chi,p}$ and $\hat{a}_{\chi,p}$ of the $p$-th collective vibration mode, so that $\hat{Q}_{\chi,p}=\sqrt{\hbar/2m\omega_{\chi,p}}(\hat{a}^{\dag}_{\chi,p}+\hat{a}_{\chi,p})$ and $\hat{P}_{\chi,p}=i\sqrt{\hbar m\omega_{\chi,p}/2}(\hat{a}^{\dag}_{\chi,p}-\hat{a}_{\chi,p})$. Then, the Hamiltonian (\ref{H0}) can be written as $\hat{H}_{0}=\sum_{\chi}\sum_{n}\hbar\omega_{\chi,n}(\hat{a}^{\dag}_{\chi,n}\hat{a}_{\chi,n}+1/2)$. Moving to rotating frame with respect to $\hat{H}_{0},$ the non-linear Hamiltonian rapidly oscillates and thus can be neglected in the vibrational rotating-wave approximation. Crucially, for a given resonance trap-frequency condition, the non-linear phonon-phonon interaction can not be neglected. For concreteness, we consider here an ion string with three ions. Then, imposing the resonance condition $\omega_{a}\approx \omega_{b}+\omega_{c},$ where $\omega_a$ is the zigzag axial normal-mode frequency and, respectively, $\omega_{b}$ and $\omega_{c}$ are the transverse rocking and zigzag mode frequencies, and neglecting the fast-rotating terms, we obtain \cite{Ding2018,Maslennikov}
\begin{equation}
\hat{H}_{\rm v}=\hbar\omega_{a}\hat{a}^{\dag}\hat{a}+\hbar\omega_{b}\hat{b}^{\dag}\hat{b}+\hbar\omega_{c}\hat{c}^{\dag}\hat{c}+\hbar\xi(\hat{a}^{\dag}\hat{b}\hat{c}+
\hat{a}\hat{b}^{\dag}\hat{c}^{\dag}),\label{nonL}
\end{equation}
where $\xi=(9\omega^{2}_{z}/5z_{0})\sqrt{\hbar/m\omega_a\omega_b\omega_c}$ is the non-linear phonon-phonon coupling and $z_{0}=(5e^{2}/16\pi\epsilon_{0}m\omega^{2}_{z})^{1/3}$ is the distance between neighbouring ions. The Hamiltonian describes processes of non-linear coherent energy exchange between the three collective vibrational modes, in which one phonon from the axial zigzag mode ($a$-mode) is converted into one phonon in the radial rocking mode ($b$-mode) and one in the radial zigzag mode ($c$-mode), and vice versa.

\section{Laser-free creation of two-mode squeezed state}\label{TM}
In the following, we propose a laser-free creation of the two-mode squeezed state, which relies on the non-linear phonon-phonon interaction, described by Hamiltonian (\ref{nonL}).

A two-mode squeezed state is defined as a coherent superposition of twin Fock states \cite{Gerry2005}
\begin{equation}
|\psi_{\rm tmss}\rangle=\hat{S}(\zeta)|0,0\rangle=\frac{1}{\cosh r}\sum_{n=0}^{\infty}(e^{i\theta}\tanh r)^{n}|n,n\rangle, \label{TMSS}
\end{equation}
where $\hat{S}(\zeta)=e^{\zeta \hat{a}^{\dag}_{A}\hat{a}^{\dag}_{B}-\zeta^{*}\hat{a}_{A}\hat{a}_{B}}$ is the two-mode squeezing operator with $\zeta=r e^{i\theta},$ where $r$ is the squeezing amplitude and $\theta$ is the phase. Here $|n,n\rangle=|n\rangle_{A}|n\rangle_{B}$ is the twin Fock state with $n$ bosons (phonons) in each mode.

Let us now assume that a time-dependent oscillating electric field is applied, therefore displaces the motional excitation of axial $a$-mode. Its effect is described by the term $\hat{H}_{F_a}(t)=F_a z_{a}\cos(\omega_d t+\varphi)(\hat{a}^{\dag}+\hat{a})$, where $F_a$ is the magnitude of the electric field, $\omega_d$ is the driving frequency, $\varphi$ is the phase, and $z_{a}=\sqrt{\hbar/2m\omega_a}$ is the spread of the zero-point wave function along the axial direction. We assume that the trap frequencies satisfy the condition
\begin{equation}
\omega_a=\omega_b+\omega_c+\omega\label{cond}
\end{equation}
and the driving frequency $\omega_d=\omega_a-\omega$ is shifted from the $a$-mode frequency $\omega_a$ by a small detuning $\omega$ ($\omega_a\gg\omega$). Moving into a rotating frame with respect to $\hat{U}_{a}(t)=e^{-i(\omega_a-\omega)t\hat{a}^{\dag}\hat{a}-i\omega_b \hat{b}^{\dag}\hat{b}-i\omega_c \hat{c}^{\dag}\hat{c}}$, the total Hamiltonian $\hat{H}=\hat{U}_{a}^{\dag}(t)(\hat{H}_{\rm v}+\hat{H}_{F_a}(t))\hat{U}_{a}(t)-i\hbar \hat{U}_{a}(t)^{\dag}\partial_{t}\hat{U}_{a}(t)$ becomes
\begin{eqnarray}
&&\hat{H}=\hbar\omega \hat{a}^{\dag}\hat{a}+\hat{H}_{a},\notag\\
&&\hat{H}_{a}=\hbar\Omega_a (\hat{a}^{\dag}e^{i\varphi}+\hat{a}e^{-i\varphi})+\hbar\xi (\hat{a}^{\dag}\hat{b}\hat{c}+
\hat{a}\hat{b}^{\dag}\hat{c}^{\dag}),\label{Hnl}
\end{eqnarray}
where $\Omega_a=F_a z_a/2\hbar$.
\begin{figure}
\includegraphics[width=0.45\textwidth]{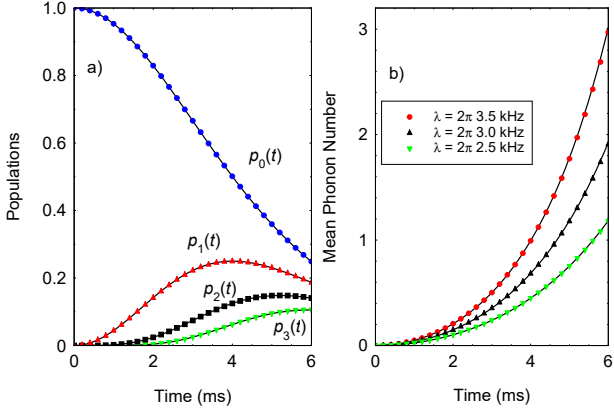}
\caption{(Color online) (a) Time evolution of the probability $p_{n}(t)$ to find the system in twin Fock state $|n,n\rangle$. We compare the probabilities derived from the original Hamiltonian (\ref{Hnl}) and the analytical expression $p_{n}(t)=\tanh^{2n}(r)/\cosh^{2}(r)$ (solid lines). We assume an initial vacuum phonon state for all three modes. The parameters are set to $\Omega_a/2\pi=3.5$ kHz, $\xi/2\pi=0.2$ kHz, and $\omega/2\pi=20$ kHz. The Hilbert space for the three modes is truncated at $n_{\rm max}=20$. (b) Mean phonon number in the radial $b$ mode for various $\lambda$ compared with the analytical result $n_{b}=\sinh^2(r)$ (solid lines).}
\label{fig1}
\end{figure}

In the following we consider a weak-coupling regime, $\Omega_a,\xi\ll\omega$, in which the phonon excitations of the $a$-mode are suppressed and thus can be eliminated from the dynamics. This can be carried out by performing the canonical transformation $\hat{U}=e^{\hat{S}_a}$ to Hamiltonian (\ref{Hnl}), such that $\hat{H}_{\rm eff}=e^{-\hat{S}_a}\hat{H}e^{\hat{S}_a}\approx \hbar\omega \hat{a}^{\dag}\hat{a}+(1/2)[\hat{H}_{a},\hat{S}_a]$ \cite{Ivanov2016} with $\hat{S}_a=(\Omega_a/\omega)(\hat{a} e^{-i\varphi}-\hat{a}^{\dag}e^{i\varphi})+(\xi/\omega)(\hat{a}\hat{b}^{\dag}\hat{c}^{\dag}-\hat{a}^{\dag}\hat{b}\hat{c})$. Keeping only the terms of order $\Omega_a\xi/\omega,$ we obtain the following effective Hamiltonian:
\begin{equation}
\hat{H}_{\rm eff}=\hbar\omega \hat{a}^{\dag}\hat{a}-\hbar\frac{\Omega_a\xi}{\omega}(e^{i\varphi}\hat{b}^{\dag}\hat{c}^{\dag}+e^{-i\varphi}\hat{b}\hat{c})+\hat{H}_a^{\prime}.\label{Heff}
\end{equation}
The results indicate that, up to leading order, the off-resonance oscillating electric force suppresses the phonon exchange between the axial $a$-mode and the other two transverse $b$- and $c$-modes. This induces an effective two-mode squeezing between the $b$- and $c$-modes with magnitude $r=(\Omega_a\xi/\omega)t$ and phase $\theta=\varphi+\pi/2$. The residual Hamiltonian is $\hat{H}_a^{\prime}=-\Omega^{2}_{a}/\omega+(\xi^{2}/\omega)(\hat{n}_{a}+\hat{n}_{a}\hat{n}_{b}+\hat{n}_{a}\hat{n}_{c}-\hat{n}_{b}\hat{n}_{c}),$ where the first term gives rise to a global phase-shift while the second term describes residual phonon-phonon interaction. It can, however, be neglected, as long as $\Omega_a\gg\xi$.
\begin{figure}
\includegraphics[width=0.45\textwidth]{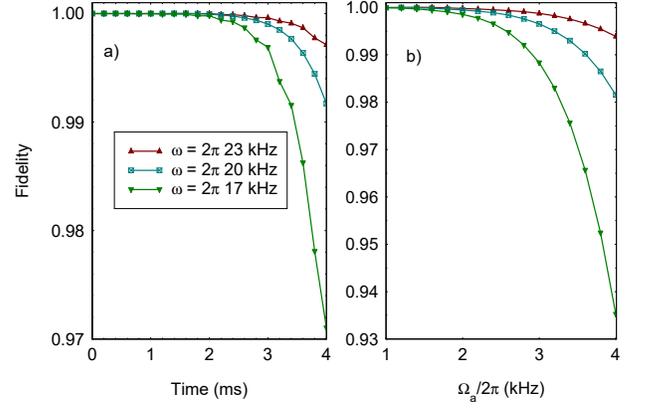}
\caption{(Color online) (a) Two-mode squeezed state fidelity, calculated from the numerical simulation of Hamiltonian (\ref{Hnl}) as a function of time for various $\omega$. The parameters are set to $\xi/2\pi=0.2$ kHz, $\Omega_a=3.5$ kHz, and $\varphi=\pi/8$. (b) Analogously but now we have set $t=4$ ms and varied $\Omega_a$ for different $\omega$.}
\label{fig2}
\end{figure}

In Fig. \ref{fig1} we compare the exact results for the twin Fock probabilities and the mean phonon number, with the analytical results, where very good agreement is observed. Furthermore, we calculate the fidelity of producing the two-mode squeezed state using $F=\langle\psi_{\rm tmss}|{\rm Tr}_{a}\{\hat{U}(t)|\psi_{\rm in}\rangle\langle\psi_{\rm in}|\hat{U}^{\dag}(t)\}|\psi_{\rm tmss}\rangle$, where a partial trace is performed over the vibrational degree of freedom of the axial $a$-mode and $\hat{U}(t)=e^{-i \hat{H} t/\hbar}$ is the unitary operator with $\hat{H}$ given by Eq. (\ref{Hnl}). The results displayed in Fig. \ref{fig2} demonstrate that, by increasing $\omega$, the higher order terms in (\ref{Heff}) can be neglected, and so the fidelity is improved. Also, for higher frequency $\Omega_{a},$ the condition for the adiabatic elimination of $a$-mode is not satisfied, which decreases the fidelity. For example, consider $\xi/2\pi=0.2$ kHz, $\Omega_a/2\pi=3.5$, $\omega/2\pi=20$, and $t_{\rm max}=4$ ms for which the infidelity estimate is of order $1-F( t_{\rm max})\approx 10^{-3}$.

\section{Beam splitter transformation}\label{BS}
We now discuss the creation of beam-splitter transformation using again the non-linear phonon-phonon interaction (\ref{nonL}). Consider now that the oscillating electric field displaces a motion amplitude either of $b$- or $c$-modes. For concreteness, assume that the displacement term is $\hat{H}_{b}(t)=F_bz_{b}\cos(\omega_{d}t+\varphi)(\hat{b}^{\dag}+\hat{b})$, where $z_{b}=\sqrt{\hbar/2m\omega_b}$, and the driving frequency is set to $\omega_d=\omega_b+\omega$ with $\omega_{b}\gg\omega$. We assume that the trap frequencies satisfy the condition (\ref{cond}). Then, moving into rotating frame with respect to $\hat{U}_{b}(t)=e^{-i\omega_a t\hat{a}^{\dag}\hat{a}-i(\omega_b+\omega) \hat{b}^{\dag}\hat{b}-i\omega_c \hat{c}^{\dag}\hat{c}},$ the total Hamiltonian becomes
\begin{eqnarray}
&&\hat{H}=-\hbar\omega \hat{b}^{\dag}\hat{b}+\hat{H}_{b},\notag\\
&&\hat{H}_{b}=\hbar\Omega_b (\hat{b}^{\dag}e^{i\varphi}+\hat{b}e^{-i\varphi})+\hbar\xi (\hat{a}^{\dag}\hat{b}\hat{c}+
\hat{a}\hat{b}^{\dag}\hat{c}^{\dag}),\label{bsH}
\end{eqnarray}
where $\Omega_b=F_b z_b/2\hbar$. Assuming that $\Omega_{b},\xi\ll\omega$ and performing a canonical transformation with $\hat{S}_{b}=(\Omega_{b}/\omega)(b^{\dag}e^{i\varphi}-\hat{b} e^{-i\varphi})+(\xi/\omega)(\hat{a}\hat{b}^{\dag}\hat{c}^{\dag}-\hat{a}^{\dag}\hat{b}\hat{c}),$ the effective Hamiltonian $\hat{H}_{\rm eff}=e^{-\hat{S}_b}\hat{H}e^{\hat{S}_b}\approx -\hbar\omega \hat{b}^{\dag}\hat{b}+(1/2)[\hat{H}_{b},\hat{S}_b]$ becomes
\begin{equation}
\hat{H}_{\rm eff}=-\hbar\omega\hat{b}^{\dag}\hat{b}+\frac{\Omega_{b}\xi}{\omega}(\hat{a}^{\dag}\hat{c} e^{i\varphi}+\hat{a}\hat{c}^{\dag}e^{-i\varphi})+\hat{H}_{b}^{\prime},
\end{equation}
where $\hat{H}_{b}^{\prime}=\Omega^{2}_{b}/\omega-(\xi^{2}/\omega)(\hat{n}_{a}+\hat{n}_{a}\hat{n}_{b}+\hat{n}_{a}\hat{n}_{c}-\hat{n}_{b}\hat{n}_{c})$. The result indicates that the off-resonant oscillating electric field suppresses the energy exchange between the $b$- mode and the other two modes, which creates a beam-splitter transformation between the $a$- and $c$-modes.

\begin{figure}
\includegraphics[width=0.45\textwidth]{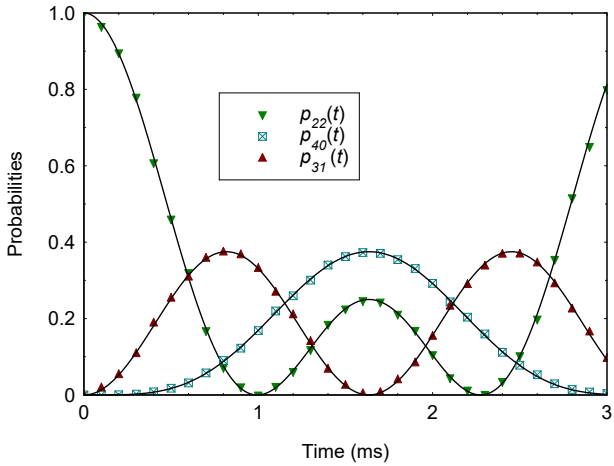}
\caption{(Color online) Probability to observe a state $|N_1,N_2\rangle$ as a function of time. We compare the probabilities derived from the original Hamiltonian (\ref{bsH}) and the analytical expressions $p_{N_1,N_2}(t)=|C^{n_1,n_2}_{N_1,N_2}|^{2}$. Due to symmetry, $p_{3,1}(t)=p_{1,2}(t)$ and $p_{4,0}(t)=p_{0,4}(t).$ We assume an initial state $|\psi_{in}\rangle=|2,0,2\rangle$. The parameters are set to $\xi/2\pi=0.2$ kHz, $\Omega_b=6.5$ kHz, and $\omega/2\pi=17$ kHz.}
\label{fig3}
\end{figure}

For arbitrary initial phonon state $|\psi_{in}\rangle=|n_1,n_2\rangle$, the beam-splitter transformation $\hat{U}_{\rm bs}(t)=e^{-i\epsilon t(\hat{a}^{\dag}\hat{c}e^{i\varphi}+\hat{a}\hat{c}^{\dag}e^{-i\varphi})}$ with $\epsilon=(\Omega_{b}\xi/\omega)$ creates a final state $|\psi_{f}\rangle=\sum_{N_1,N_{2}}e^{-i(\varphi-\frac{\pi}{2})(n_1-N_1)}C^{n_1,n_2}_{N_1,N_2}|N_1,N_2\rangle$, where, \cite{Kim2002}:
\begin{eqnarray}
C^{n_1,n_2}_{N_1,N_2}\!\!\!&=&\!\!\!\sum_{k=0}^{n_1}\sum_{l=0}^{n_2}e^{i\pi(n_1-k)}\sin^{n_1+n_2-k-l}(\epsilon t)\cos^{k+l}(\epsilon t)\notag\\
&&\!\!\!\times\frac{\sqrt{n_1!n_2!N_1!N_2!}}{k!(n_1-k)!l!(n_2-l)!}\delta_{N_1,n_2+k-l}\delta_{N_2,n_1-k+l}.
\end{eqnarray}
In Fig. (\ref{fig3}) we display the exact results for the probabilities to observe $N_1$ phonons in $a$-mode and $N_2$ phonons in $c$-mode, compared to the exact result $p_{N_1,N_2}(t)=|C^{n_1,n_2}_{N_1,N_2}|^{2}$. We see that the phonon dynamics of $b$-mode are suppressed due to the time-varying oscillating force, which causes an effective beam-splitter transformation between the other two vibrational modes.
\section{Quantum Metrology}\label{QM}
In the following section we discuss the metrological properties for parameter estimation of the two-mode squeezed state and the beam-splitter transformation.
\subsection{Generalized theoretical framework for single-parameter quantum estimation}
The value of $\lambda$, the parameter we wish to estimate, can be extracted by performing suitable discrete set of measurements, defined in terms of its corresponding positive-operator-valued measurement (POVM), $\{\hat{\Pi}_{n}\}$, with $\sum_{n}\hat{\Pi}_{n}=1$. The corresponding classical Fisher information (CFI), quantifying the amount of information on the unknown parameter, is given by \cite{Paris2009}
\begin{equation}
\mathcal{F}_{\rm C}(\lambda)=\sum_{n}\frac{1}{p_{n}(\lambda)}\left(\frac{d p_{n}(\lambda)}{d\lambda}\right)^{2},
\end{equation}
where $p_{n}(\lambda)=\langle\psi_{\lambda}|\hat{\Pi}_{n}|\psi_{\lambda}\rangle$ is the probability to get outcome $n$ from the performed measurement, and $|\psi_{\lambda}\rangle$ is the state vector of the system and depends on the parameter $\lambda$. The statistical uncertainty in the parameter estimation is quantified by the Cram\'er-Rao bound:
\begin{equation}
\delta\lambda^{2}\ge \frac{1}{\mathcal{F}_{\rm C}(\lambda)}.
\end{equation}
The statistical uncertainty however depends on the choice of an observable. Therefore, optimizing over all possible measurements allows us to determine the unknown parameter with an ultimate precision. Indeed, the CFI is upper-bounded by $\mathcal{F}_{\rm C}(\lambda)\leq \mathcal{F}_{\rm Q}(\lambda)$, where $\mathcal{F}_{\rm Q}(\lambda)={\rm Tr}(\hat{\rho}_{\lambda}\hat{L}^{2})$ is the quantum Fisher information (QFI), with $\hat{L}$ being the so-called symmetrical logarithmic derivative (SLD) operator, which satisfies the operator equation $2\partial_{\lambda}\hat{\rho}_{\lambda}=\hat{\rho}_{\lambda}\hat{L}+\hat{L}\hat{\rho}_{\lambda}$. For pure states the QFI is $\mathcal{F}_{\rm Q}(\lambda)=4(\langle\partial_{\lambda}\psi_{\lambda}|\partial_{\lambda}\psi_{\lambda}\rangle-|\langle\psi_{\lambda}|\partial_{\lambda}\psi_{\lambda}\rangle|^{2})$. Furthermore, if $|\psi_{\lambda}\rangle=e^{i\lambda \hat{G}}|\psi_{0}\rangle$, the QFI is equal to four times the variance of $\hat{G}$ with respect to the initial state, namely $\mathcal{F}_{\rm Q}(\lambda)=4\Delta \hat{G}^{2}=4(\langle \hat{G}^{2}\rangle-\langle \hat{G}\rangle^{2})$. Finally, the ultimate precision, which can be achieved, is quantified by the quantum Cram\'er-Rao bound
\begin{equation}
\delta\lambda^{2}\geq\frac{1}{\mathcal{F}_{\rm Q}(\lambda)}.
\end{equation}
For a single parameter estimation there always exists a measurement basis in which CFI is equal to QFI. Such a basis is formed by the eigenstates of the SLD operator. In general, a set of measurements that lead to the equality $\mathcal{F}_{\rm C}(\lambda)=\mathcal{F}_{\rm Q}(\lambda)$ is called optimal.
\begin{figure}
\includegraphics[width=0.45\textwidth]{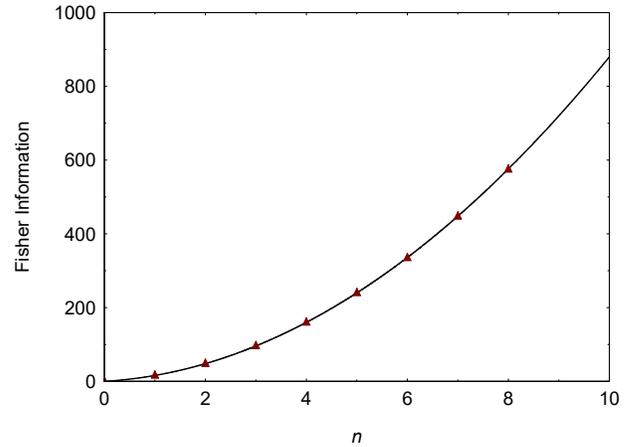}
\caption{(Color online) Classical Fisher information for the probability $p_{k}(t_f)=\langle k|{\rm Tr}_{a,b}\{|\psi(t)\rangle\langle\psi(t)|\}|k\rangle$ to observe $k$ phonons in $c$-mode as a function of the initial number of phonons $n$. The numerical result with Hamiltonian (\ref{bsH}) is compared with the QFI (\ref{qfi1}) (solid line). The parameters are set $t_{f}=1$ ms, $\Omega_b/2\pi=4.5$ kHz, $\omega/2\pi=20$ kHz, and $\xi/2\pi=0.2$ kHz.}
\label{fig4}
\end{figure}

For a quantum state with $n$ uncorrelated particles, the statistical uncertainty obeys the scaling $\delta\lambda^{2}\sim 1/n$ which is the shot-noise limit. The super-resolution and the so-called Heisenberg sensitivity can be achieved by using entangled states where $\delta\lambda^{2}\sim 1/n^{2}$ \cite{Degen2017,Pezze2018,Giovannetti2011,Giovannetti2006}.

\subsection{Two-mode squeezed state}
Consider that the parameter we wish to estimate is the magnitude of the two-mode squeezing operator, namely $\lambda=r$. Then, it is straightforward to show that the QFI is $\mathcal{F}_{\rm Q}(r)=4$. We find that the optimal basis consists of measurements of Fock state probabilities of one of the vibrational modes $p_{n}=\tanh^{2n}(r)/\cosh^{2}(r)$ which leads to equality $\mathcal{F}_{\rm C}(r)=\mathcal{F}_{\rm Q}(r)=4$. We may also consider an estimation of the phase $\theta$ in Eq. (\ref{TMSS}). Then, we have $\mathcal{F}_{\rm Q}(\theta)=4 \bar{n}(\bar{n}+1)$, where $\bar{n}=\sinh^{2}(r)$ is the mean-phonon number of one of the vibrational modes. Note that the Fock state probabilities are independent on $\theta$ so that the optimal basis is given by the eigenvectors of the SLD operator.

\subsection{Heisenberg precision }
We now discuss the quantum metrology using beam-splitter transformation. We assume that the system is prepared initially in the Fock state with $n$ phonons in each mode, $|\psi_{in}\rangle=|n,n\rangle$. Then, the system evolves into the final state $|\psi_{f}\rangle=e^{-i\epsilon t(\hat{a}^{\dag}\hat{c}+\hat{a}\hat{c}^{\dag})}|n,n\rangle$. Consider that the parameter we wish to estimate is $\lambda=\epsilon$, then the QFI information is given by
\begin{equation}
\mathcal{F}_{\rm Q}(\epsilon)=8 n(n+1),\label{qfi1}
\end{equation}
which gives Heisenberg-limited precision. Note that, because $(n^{2}+n)^{-1}<n^{-2},$ a sub-Heisenberg limit in the parameter estimation can be achieved \cite{Anisimov2010}.

In Fig. \ref{fig4} we show the exact result for the CFI, using the full Hamiltonian (\ref{bsH}) which is compared to the QFI (\ref{qfi1}). We assume that the parameter $\epsilon$ is extracted by measuring the probabilities to observe a Fock state with $k$ phonons in one of the vibrational modes, namely $p_{k}(t)=\langle k|{\rm Tr}_{a,b}\{|\psi(t)\rangle\langle\psi(t)|\}|k\rangle$, where $|\psi(t)\rangle$ is the exact state vector of the total system. Crucially, the measurement of the Fock state probabilities is an optimal, in sense that the CFI is equal to the QFI, as can be seen from Fig. \ref{fig4}. Finally, we point out that, recently, an efficient adiabatic technique for measuring the Fock state probabilities with trapped ions was proposed \cite{Kirkova2021}.
\begin{figure}
\includegraphics[width=0.45\textwidth]{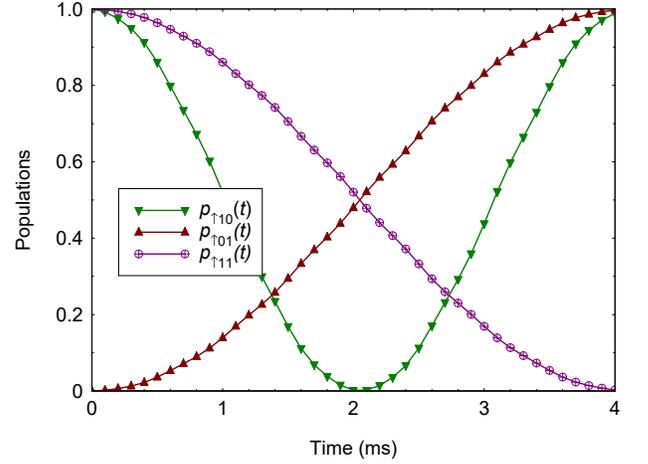}
\caption{(Color online) Conditional beam-splitter transformation. The system is prepared initially in the state $\left|\uparrow,1,0\right\rangle$ where the motional $a$- and $c$-modes are swapped at the gate time $t_g$ to $\left|\uparrow,0,1\right\rangle$. We show the respective probabilities $p_{\uparrow,1,0}(t)$ and $p_{\uparrow,0,1}(t)$ which are derived from the original Hamiltonian (\ref{laserion}). The probability $p_{\uparrow,1,1}(t)$ when the system is prepared initially in the state $\left|\uparrow,1,1\right\rangle$. The parameters are set to $\xi/2\pi=0.2$ kHz, $g_b/2\pi=5.5$ kHz, and $\omega/2\pi=18$ kHz, and $\eta_{b}=0.06$.}
\label{fig5}
\end{figure}

\section{Conditional Beam-Splitter Transformation}\label{CBST}
In this Section we extend the discussion to the creation of spin-dependent (conditional) beam-splitter transformation. Let us consider the ion's two-metastable levels $\left|\uparrow\right\rangle$ and $\left|\downarrow\right\rangle$ which compose an effective spin system with transition frequency $\omega_0$. Instead of applying an oscillating electric field, one can decouple the $b$-mode from the other two $a$- and $c$- modes using a laser field onto a single ion.
Assume that the first ion is addressed by pairs of laser beams in a Raman configuration with a wave-vector difference $\Delta \vec{k}$ along the transverse $x$-direction ($|\Delta \vec{k}|=k_x$) and laser frequency difference $\Delta\omega_{\rm L}=\omega_b+\omega$ tuned near the $b$-mode. Such a laser configuration generates a spin-dependent force, which provides a coupling between the spin state $\left|\uparrow\right\rangle$ and the $b$-mode \cite{Lee2005}. The Hamiltonian describing the laser-ion interaction is given by $\hat{H}_{I}(t)=\hbar\Omega\{e^{ik_x\delta \hat{r}_{x}-\Delta\omega_{\rm L}t}+{\rm H.c.}\}\left|\uparrow\right\rangle\left\langle\uparrow\right|$, where $\Omega$ is the Rabi frequency \cite{Lee2005,Wineland1998,Haffner2008,Schneider2012}. Let us define the Lamb-Dicke parameter $\eta_{b}=k_{x}\sqrt{\frac{\hbar}{2m\omega_{b}}}M_{1,b}^{x}$. Then, expanding the laser-ion interaction in power series of $\eta_{b}$ and neglecting the fast rotating terms, we obtain
\begin{eqnarray}
\hat{H}&=&-\hbar\omega\hat{b}^{\dag}\hat{b}+\hbar g_{b}(\hat{b}^{\dag}\hat{F}(\hat{n}_{b})+\hat{F}(\hat{n}_{b})\hat{b})\left|\uparrow\right\rangle\left\langle\uparrow\right|
\notag\\
&&+\hbar\xi(\hat{a}^{\dag}\hat{b}\hat{c}+\hat{a}\hat{b}^{\dag}\hat{c}^{\dag}),\label{laserion}
\end{eqnarray}
where $g_{b}=\Omega\eta_{b}$ is the spin-phonon coupling. The non-linear operator, which quantifies the higher-order terms in the Lamb-Dicke expansion, is given by \cite{Vogel1995}
\begin{equation}
\hat{F}(\hat{n}_{b})=e^{-\eta^{2}_{b}/2}\sum_{n=0}^{\infty}\frac{(-\eta_{b}^{2})^n}{n!(n+1)!}\hat{b}^{\dag n}\hat{b}^{n}.
\end{equation}
Note that, for sufficiently strong coupling $g_{b},$ but still $g_{b}\ll\omega,$ the coupling strength of the higher-order terms in the Lamb-Dicke expansion is compatible with the non-linear coupling $\xi$. In order to account for this, we include these terms in the numerical simulations.
\begin{figure}
\includegraphics[width=0.45\textwidth]{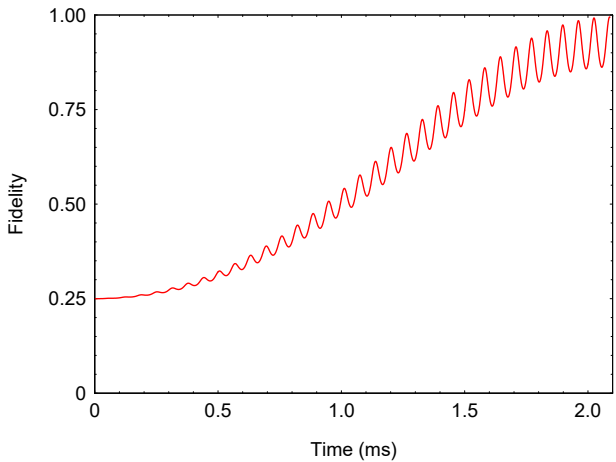}
\caption{(Color online) Fidelity for the creation of spin-dependent N00N state as a function of time for $n=2$, calculated from the numerical simulation with Hamiltonian (\ref{laserion}). We assume that an additional term $\hat{H}_{\rm AC}=-(g^{2}/\omega)\left|\uparrow\right\rangle\left\langle\right\uparrow|$ is applied to (\ref{laserion}), and compensates the shift of the spin frequency due to the spin-motional interaction. The parameters are set to $g_b/2\pi=6.3$ kHz, and $\omega/2\pi=15.8$ kHz, $\xi/2\pi=0.3$ kHz and $\eta_{b}=0.05$}
\label{fig6}
\end{figure}

Here, as an example, we study the efficiency of generation of Fredkin gate, where the control qubit consists of the two internal states of the ion and the target qubits are composed of the Fock states of $a$- and $c$-modes. We have
\begin{equation}
\hat{U}_{F}\left|\downarrow,n,m\right\rangle=\left|\downarrow,n,m\right\rangle,\quad \hat{U}_{F}\left|\uparrow,n,m\right\rangle=(-i)^{n+m}\left|\uparrow,m,n\right\rangle,\label{fredkin}
\end{equation}
where $\hat{U}_{F}$ is the Fredkin gate. In the following we show that, using a laser-ion interaction and the non-linear phonon-phonon interaction (\ref{laserion}), we can implement the Fredkin gate. \\
Ideally, the Fredkin gate can be carried out using the conditional beam-splitter transformation $\hat{U}_{\rm cbs}(t)=e^{-i\epsilon_{b} t\left|\uparrow\right\rangle\left\langle\uparrow\right|(\hat{a}^{\dag}\hat{c}+\hat{a}\hat{c}^{\dag})}$ with $\epsilon_{b}=g_{b}\xi/\omega$ such that, for time $t_g=\frac{\pi}{2\epsilon_{b}}$, we have $\hat{U}_{F}=\hat{U}_{\rm sbm}(t_g)$.

In Fig. \ref{fig5} we show the exact result for the probabilities to observe states $\left|\uparrow,1,0\right\rangle$ and $\left|\uparrow,0,1\right\rangle$ when the system is prepared initially in the state $\left|\uparrow,1,0\right\rangle$. We also show $p_{\uparrow,1,1}(t)$ when the system is prepared in the state $\left|\uparrow,1,1\right\rangle$. At the gate time $t_g$, the two motional $a$- and $c$- modes are swapped only if the spin is in the excited state $\left|\uparrow\right\rangle$, according to (\ref{fredkin}).

Furthermore, we consider the generation of a spin-dependent N00N-like state $\left|\psi_{\rm N00N}\right\rangle=(\left|\downarrow\right\rangle\left|n,0\right\rangle+(-i)^{n}\left|\uparrow\right\rangle\left|0,n\right\rangle)/\sqrt{2}$, using the Fredkin gate (\ref{fredkin}). We assume that the system is initially prepared in the state $\left|\psi_{in}\right\rangle=|\psi_{\rm spin}\rangle|n,0\rangle$ with $|\psi_{\rm spin}\rangle=(\left|\downarrow\right\rangle+\left|\uparrow\right\rangle)/\sqrt{2}$, such that, after the application of the Fredkin gate, we create the entangled state $\left|\psi_{\rm N00N}\right\rangle$. As a figure of merit for the fidelity, we consider $F_{G}(t)=\langle\psi_{\rm N00N}|{\rm Tr}_{b}\{\hat{U}(t)|\psi_{in}\rangle\langle\psi_{ in}|\hat{U}^{\dag}(t)\}|\psi_{\rm N00N}\rangle$, where a partial trace is performed over the vibrational degree of freedom of the radial $b$-mode, and $\hat{U}(t)=e^{-i \hat{H} t/\hbar}$ is the unitary operator, with $\hat{H}$ given by Eq. (\ref{laserion}). In Fig. \ref{fig6} we display the numerical result for $F_{G}(t)$. We assume that the shift of the spin frequency, due to the spin-phonon interaction, is compensated by the term $\hat{H}_{\rm AC}=-(g^{2}/\omega)\left|\uparrow\right\rangle\left\langle\right\uparrow|$. For example, consider the parameters $g_b/2\pi=6.3$ kHz, and $\omega/2\pi=15.8$ kHz, for which we estimate infidelity of order of $1-F_{G}(t_g)\sim 6\times 10^{-3}$.

\section{Conclusion}\label{C}

In conclusion, we have proposed a laser-free method for creation of two-mode squeezed states which relies on the coupling between the axial zigzag mode and the radial rocking and zigzag modes in a linear ion crystal. We have shown that a weak, time-varying electric field suppresses the energy exchange between the axial mode and the other two radial modes, and leads to an effective two-mode squeezing of the $b$- and $c$- modes. The same technique can be used for the creation of a beam-splitter transformation between the axial $a$-mode and one of the radial $b$- or $c$-modes. We have studied the quantum metrological properties of the two-mode squeezed state and the beam-splitter transformation. We have shown that, for an initial state with $n$-phonons in each mode, one can achieve sub-Heisenberg limit of precision using a beam-splitter transformation, where the quantum Fisher information scales with the number of phonons as $n^{2}+n$. Finally, we have discussed the implementation of a conditional beam-splitter-transformation. In that case a spin-dependent force is applied to one of the radial modes. We have shown that the combination of the spin-dependent force and phonon non-linearity leads to the creation of conditional beam-splitter transformation. For a given gate time, the conditional beam-splitter transformation is identical to the three-qubit Fredkin gate, in which the target qubit consists of the internal states of the ion and the control qubits are the phonon Fock states. Finally, we have shown that the Fredkin gate can be used for the creation of high-fidelity quantum N00N states.

%%%%%%%%%%%%%%%%%%%%%%%
\section*{Acknowledgments}
This research is supported by the Bulgarian national plan for recovery and resilience, contract BG-RRP-2.004-0008-C01 (SUMMIT: Sofia University Marking Momentum for Innovation and Technological Transfer), project number 3.1.4.

\appendix
%%%%%%%%%%%%%%%%%%%%%%%%%%%%%%%%%
%%%%%%%%%%%%%%%%%%%%%%%%%%%%%%%%%

\end{document}